\documentclass[aps,prl,twocolumn,groupedaddress]{revtex4}

\usepackage{graphicx}

\begin{document}

\def\eps{\varepsilon}
\def\ms{\overline{\rm MS}}
\def\lms{\Lambda_{\ms}}
\def\disg{{\rm DIS}_\gamma}

\def\f2y{F_2^\gamma}
\def\fqy{f_{q/\gamma}}
\def\fby{f_{\bar{q}/\gamma}}
\def\fgy{f_{g/\gamma}}
\def\d  {{\rm d}}
\def\O  {{\cal O}}

\def\lp{\left. }
\def\rp{\right. }
\def\lr{\left( }
\def\rr{\right) }
\def\le{\left[ }
\def\re{\right] }
\def\lg{\left\{ }
\def\rg{\right\} }
\def\lb{\left| }
\def\rb{\right| }

\def\beq{\begin{equation}}
\def\eeq{\end{equation}}
\def\bea{\begin{eqnarray}}
\def\eea{\end{eqnarray}}

\preprint{DESY 02-052}
\preprint{FERMILAB-Pub-02/071-E}
\title{Strong Coupling Constant from the Photon Structure Function}
\author{Simon Albino}
\affiliation{{II.} Institut f\"ur Theoretische Physik, Universit\"at Hamburg,
             Luruper Chaussee 149, D-22761 Hamburg, Germany}
\author{Michael Klasen}
\email[]{michael.klasen@desy.de}
\affiliation{{II.} Institut f\"ur Theoretische Physik, Universit\"at Hamburg,
             Luruper Chaussee 149, D-22761 Hamburg, Germany}
\author{Stefan S\"oldner-Rembold}
\thanks{Heisenberg Fellow}
\affiliation{FNAL, P.O.\ Box 500, MS 357, Batavia, IL 60510, USA}
\date{\today}
\begin{abstract}
We extract the value of the strong coupling constant $\alpha_s$ from a
single-parameter pointlike fit to the photon structure function $\f2y$ at large
$x$ and $Q^2$ and from a first five-parameter full (pointlike and hadronic) fit
to the complete $\f2y$ data set taken at PETRA, TRISTAN, and LEP. In
next-to-leading order and the $\ms$ renormalization and factorization schemes,
we obtain
	$\alpha_s(m_Z)=0.1183\pm0.0050$(exp.)$^{+0.0029}_{-0.0028}$(theor.)
	[pointlike]
and 
	$\alpha_s(m_Z)=0.1198\pm0.0028$(exp.)$^{+0.0034}_{-0.0046}$(theor.)
	[pointlike and hadronic].
We demonstrate that the data taken at LEP have reduced the experimental error
by about a factor of two, so that a competitive determination of $\alpha_s$
from $\f2y$ is now possible.
\end{abstract}
\maketitle


The theory of strong interactions, Quantum Chromodynamics (QCD), is one of the
corner stones of the Standard Model of elementary particle physics. The
precise determination of its fundamental parameter, the strong coupling
constant $\alpha_s$, bears important implications for the validity not only
of QCD itself, but also of even more fundamental theories, since these have to
contain the Standard Model as an effective field theory in the low-energy
limit. A more fundamental theory, which might explain the size of the strong
coupling constant, has yet to be established. Therefore, $\alpha_s$ must
currently be extracted from experiment. Among the large variety of processes
that have been used to this end, the most precise values have been obtained in
$Z$-boson- and $\tau$-decays at LEP, scaling violations in structure
functions at HERA, and quarkonium decay branching fractions and lattice
calculations of quarkonium mass splittings, leading -- together with other,
less precise measurements -- to a current world average of $\alpha_s(m_Z)=
0.1172\pm 0.0020$ at the mass of the $Z$-boson, $m_Z=91.1876$ GeV
\cite{Groom:2000in}.

When the photon structure function $\f2y$ was first discussed in the context of
QCD, a precise determination of $\alpha_s$ quickly emerged as one of its most
interesting applications. Due to the pointlike coupling of the photon to
quarks, the leading order (LO, $\O(\alpha/\alpha_s)$) \cite{Witten:1977ju} and
next-to-leading order (NLO, $\O(\alpha)$) \cite{Bardeen:1979hg} contributions
to
$\f2y(x,Q^2)$ are calculable in QCD perturbation theory, if the virtuality $Q$
in the deep-inelastic electron-photon scattering process is significantly
larger than the asymptotic scale parameter $\Lambda$. Unfortunately, this
pointlike contribution exhibits a power singularity at small
Bjorken-$x$ \cite{Duke:1980ij}, which becomes rapidly stronger in higher orders
\cite{Rossi:1983bp}. The singularity can be regularized with a non-perturbative
\cite{Antoniadis:1983fv} or transverse-momentum \cite{Field:1986gf} cut-off,
but then the sensitivity to $\alpha_s$ is reduced and a dependence on the
unphysical cut-off is introduced \cite{Frazer:1987sb}. It is then necessary to
fit both $\alpha_s$ and the cut-off to experimental data. Alternatively, the
singularity can be canceled order by order in perturbation theory by retaining
a hadronic boundary condition at a low starting scale $Q_0$
\cite{Gluck:1983mm}.
In this case it is necessary to fit $\alpha_s$ and the hadronic
input to experimental data. However, the evolution of the hadronic input to the
physical scale $Q$ is still predicted by perturbative QCD through inhomogeneous
evolution equations \cite{DeWitt:1979wn}, and the negligibility of the hadronic
input can be tested {\em a posteriori}. Both methods have been applied in the
past to PEP and PETRA data yielding $\Lambda^{(4)}_{\ms}=
180^{+100}_{-~90}$ MeV \cite{Wagner:1986dj} or $\alpha_s(m_Z)=0.108
^{+0.008}_{-0.010}$. This value contributed to the world average in the 1988
\cite{Yost:1988ke}, 1990 \cite{Hernandez:1990yc}, and 1992 \cite{Hikasa:1992je}
issues of the Review of Particle Properties, but was then abandoned on the
grounds that there were ``no new results and the data do not contribute
significantly to the average'' \cite{Montanet:1994xu}. Since then it has been
widely believed \cite{Gluck:1992jc,Aurenche:1994in,Schuler:1995fk,
Gordon:1997pm,Gluck:1999ub} that the sensitivity of $\f2y$ to $\alpha_s$ is
small.

In this Letter, we point out that over the last decade a wealth of new
$\f2y$ data has been collected at the $e^+e^-$-colliders TRISTAN and LEP, which
extends to high average values of $Q^2$, $\langle Q^2\rangle\leq 780$
GeV$^2$. We demonstrate that the new data improve the sensitivity of
$\f2y$ to $\alpha_s$ significantly and that a single-parameter pointlike fit
as well as a five-parameter full (pointlike and hadronic) fit to PETRA,
TRISTAN, and LEP data yield results, which are not only consistent with the
world average, but also have competitive experimental and theoretical errors.

We work in a fixed flavor number scheme with three active quark flavors ($u,d,
s$). It is well known \cite{Gluck:1992ee} that for current measurements of
$\f2y$ the available hadronic energy squared $W^2=Q^2(1-x)/x$ is not much
larger than the production threshold $4m_h^2$ of the heavy quarks ($h=c,b,t$),
so that mass effects can not be neglected and the massive, fixed order
$\O(\alpha)$ expression for the Bethe-Heitler process $\gamma^\ast(Q^2)\gamma
\to h\bar{h}$ \cite{Budnev:1974de} should be used instead of the massless,
factorized $\O(\alpha/\alpha_s)$ expression. For a consistent NLO
($\O(\alpha)$) analysis, we
do not include the known \cite{Laenen:1994ce}, but numerically small,
$\O(\alpha\alpha_s)$ corrections to the Bethe-Heitler process and omit the
$\O(\alpha\alpha_s^2)$ contributions from the process $\gamma^\ast(Q^2)g\to
h\bar{h}$. The heavy quark masses are not well constrained from measurements
of $\f2y$. We adopt a charm quark mass of $m_c=1.5\pm0.1$ GeV in good agreement
with recent precise determinations from threshold production at
$e^+e^-$-colliders \cite{Kuhn:2001dm}. At large $Q^2$, charm quarks contribute
up to 40 \% to $\f2y$ in the whole $x$-range, while at small $Q^2$ they
contribute at most 10\% below $x\sim 0.2$. The contribution from bottom quarks
is suppressed by a relative factor $1/16$ from the different quark charges,
while the threshold for top quark production lies at extremely small
$x=10^{-5}...10^{-3}$, so that these contributions are both numerically
negligible.

Since we wish to omit spurious higher order terms, which arise from the
convolution of NLO contributions to the parton densities with the NLO Wilson
coefficients and lead to instabilities at large $x$ \cite{Gluck:1992ee},
we work in
Mellin moment space, where the convolutions reduce to simple products, the
evolution can be done analytically and without any approximations, and
spurious higher order terms can be consistently omitted. The resulting
prediction for $\f2y$ is then converted back to $x$-space using a fast inverse
Mellin transform with logarithmic mapping and fitted to experimental
measurements with the multidimensional minimization algorithm MINUIT
\cite{James:1975dr}. The quality of the fit is measured in terms of the
$\chi^2$ value per degree of freedom, $\chi^2$/DF, for all selected data
points.

We include in our analysis all published measurements of $\f2y$ collected at
the high-energy $e^+e^-$-colliders PETRA \cite{Bartel:1984cg,Berger:1984xt,
Althoff:1986fi}, TRISTAN \cite{Sahu:1995gj,Kojima:1997wg,
Muramatsu:1994rq}, and LEP \cite{Barate:1999qy,Abreu:1996xt,Acciarri:1998ig,
Ackerstaff:1997ni,Ackerstaff:1997se,
Abbiendi:2000cw,Abbiendi:2002te}. If more than one set of statistically
overlapping data exists, the most recent publication is used. We exclude from
our fit the data published by the TPC/2$\gamma$ Collaboration at PEP
\cite{Aihara:1987xq,Aihara:1987xw}, since several data points, mainly at low
$x$, are inconsistent with measurements
published by PLUTO \cite{Berger:1984xt}, L3 \cite{Acciarri:1998ig}, and OPAL
\cite{Abbiendi:2000cw} in the range $1.9 < Q^2 < 5.1$~GeV$^2$. Data where the
charm component has been subtracted are also discarded. Statistical
uncertainties and correlations between data points due to the experimental
unfolding are taken into account as provided by the experiments, while
systematic uncertainties are assumed to be uncorrelated. Due to this assumption
the values of $\chi^2$/DF are expected to be on average slightly less than
unity. If asymmetric errors
are given by the experiments, the data points are taken at the center of the
full error interval. Most experiments have not corrected for the finite
virtuality of the target photon $P^2$. We neglect $P^2$ in this analysis,
since usually $P^2\ll Q^2$.

For our pointlike fit, we identify the starting scale $Q_0$ with the asymptotic
scale parameter $\Lambda$, so that the hadronic input vanishes automatically
and only a single parameter ($\Lambda$, or equivalently $\alpha_s(m_Z)$) has to
be fitted. As discussed above, this is only justified at large $x$ and $Q^2$,
where the residue of the pointlike singularity is expected to be small.
Therefore, we perform our single-parameter pointlike fit only to a subset of
data points with $x\geq 0.45$ and $Q^2\geq59$ GeV$^2$. Very similar results
are obtained with the widely used values of $Q_0=0.5...0.6$ GeV
\cite{Gluck:1992jc,Aurenche:1994in,Schuler:1995fk, Gluck:1999ub}, while
choosing $Q_0=1$ GeV significantly increases the value of $\chi^2$/DF;
two-parameter pointlike fits of $\alpha_s$ and $Q_0$ are driven to
$Q_0\simeq\Lambda$. In the first three lines of Tab.\ \ref{tab:1}
%
\begin{table}
\caption{\label{tab:1}$\chi^2$/DF and $\alpha_s(m_Z)$ values obtained in LO and
         NLO in the $\ms$ and $\disg$ factorization schemes with a
         single-parameter fit of
         the pointlike photon structure function $\f2y$. Also shown are the 
         results obtained without LEP data and with very high $Q^2$ data.}
\begin{ruledtabular}
\begin{tabular}{llc}
       Scheme & ~~~$\chi^2/$DF& $\alpha_s(m_Z)$ \\
\hline
       LO     & ~~7.9/~19& $0.1260\pm0.0055$(ex)$^{+0.0061}_{-0.0055}$(th) \\
       $\ms$  & ~~9.1/~19& $0.1183\pm0.0050$(ex)$^{+0.0029}_{-0.0028}$(th) \\
       $\disg$& ~~8.1/~19& $0.1195\pm0.0051$(ex)$^{+0.0031}_{-0.0028}$(th) \\
\hline
        w/o LEP &~~3.2/~~7& $0.1244\pm0.0126$(ex)$^{+0.0033}_{-0.0032}$(th) \\
        high $Q^2$ &~11.9/~~8& $0.1159\pm0.0125$(ex)$^{+0.0018}_{-0.0018}$(th) \\
\end{tabular}
\end{ruledtabular}
\end{table}
%
we list the $\chi^2$/DF and $\alpha_s(m_Z)$ values obtained in LO and NLO.
The NLO fit is performed in two factorization schemes ($\ms$ and
$\disg$ \cite{Gluck:1992ee}) with different treatment of the pointlike
Wilson coefficient in $\f2y$, but the numerical variation is found to be
small. The total values of $\chi^2$/DF as well as those for the
individual data sets (not shown) lie around unity or below, indicating that
the pointlike photon structure function and the fitted values of
$\alpha_s(m_Z)$ describe the data sets well within their statistical and
systematic uncertainties. The experimental errors are determined by varying
$\alpha_s(m_Z)$ until the total value of $\chi^2$ is increased by one unit.
To estimate the theoretical error, we vary the charm quark mass as indicated
above and follow the common convention of varying the factorization and
renormalization scales by factors of two about their central value, the
physical scale $Q$. We then add these three individual errors in quadrature.
The LO value of $\alpha_s(m_Z)$ is consistent
with the NLO value within the expected accuracy, $\O(\alpha_s^2)$, and the
theoretical error is reduced from LO to NLO as expected.
In the fourth line of Tab.\ \ref{tab:1}, we list the result of a fit without
the LEP data. The experimental error is more than doubled, showing that the
LEP data have considerably increased the sensitivity of $\f2y$ to $\alpha_s$
at high $x$ and $Q^2$.
When data at all values of $x$, but very high $Q^2$ ($Q^2\geq 284$ GeV$^2$)
are fitted, the central value of $\alpha_s(m_Z)$ remains virtually unchanged
(last line of Tab.\ \ref{tab:1}).
At very high $Q^2$, the
theoretical error drops by a factor of two, whereas the experimental error
increases. Measurements of $\f2y$ at a future linear $e^+e^-$- or
$e\gamma$-collider like TESLA at very high values of $Q^2$ and with small
experimental errors will therefore lead to even more precise determinations of
$\alpha_s$.

\begin{figure}
 \centering
 \includegraphics[width=0.9\columnwidth]{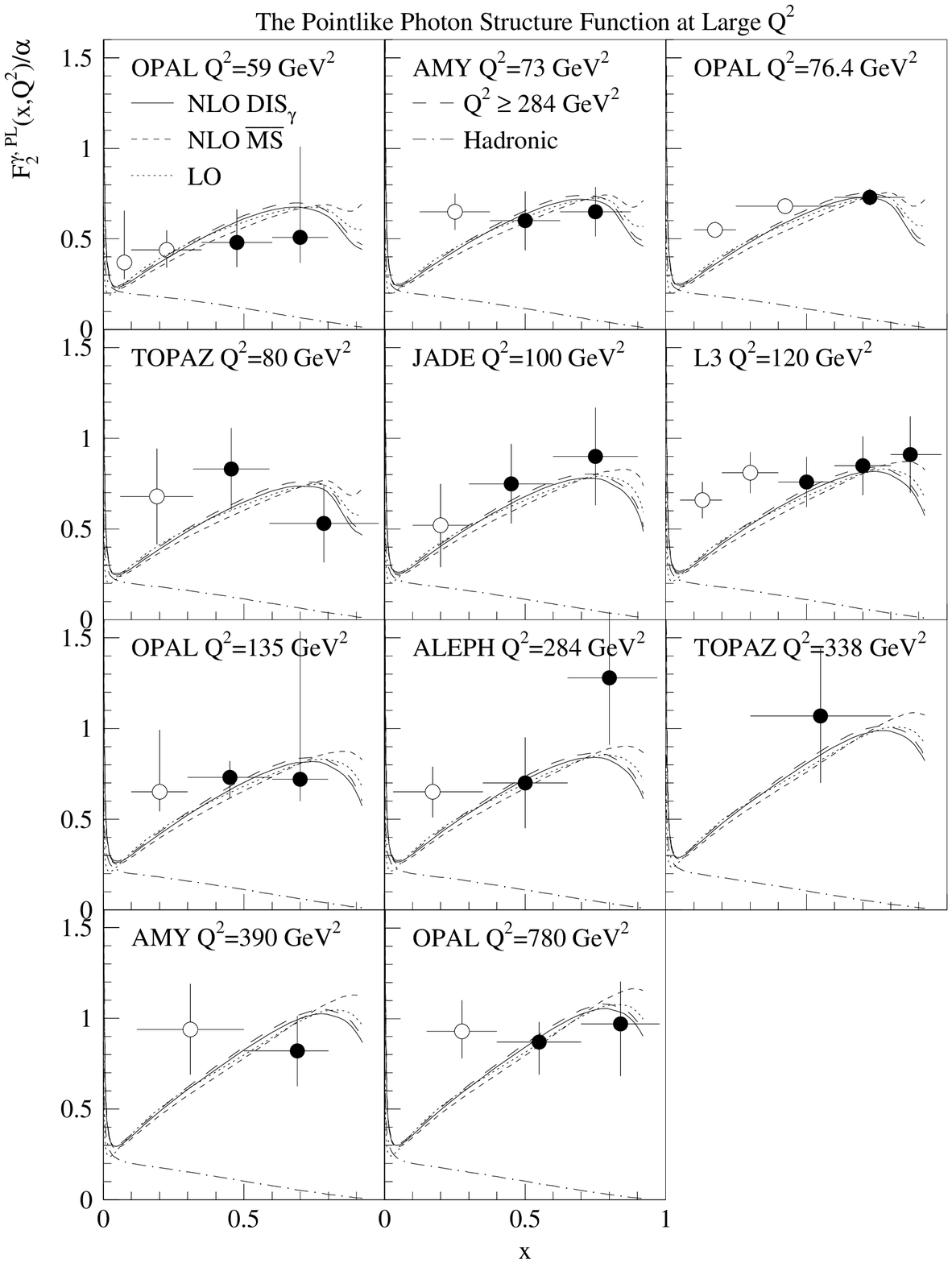}
 \caption{\label{fig:1}Single-parameter fits of the pointlike photon structure
 function, compared to PETRA \cite{Bartel:1984cg}, TRISTAN \cite{Sahu:1995gj,
 Muramatsu:1994rq}, and LEP \cite{Barate:1999qy,Acciarri:1998ig,
 Ackerstaff:1997ni,Ackerstaff:1997se,Abbiendi:2002te} data at large
 $Q^2$. The data points marked by open circles have not been used in
 the fits. Also shown is the hadronic contribution from a five-parameter NLO
 fit of the full photon structure function in the $\disg$ scheme.}
\end{figure}

The goodness of our pointlike fit may also be judged from Fig.\ \ref{fig:1},
%
%
where the fitted data points are shown as full circles, while those that have
been omitted from the fit are shown as open circles, and where the statistical
and systematic errors have been added in quadrature. The theoretical curves
are perturbatively stable, {\it i.e.} LO and NLO fits differ only by small
amounts. The choice of factorization scheme affects the region of very
large $x$, but it has only a minor effect on the description of the data. 
Also shown in Fig.\ \ref{fig:1} is the hadronic contribution from a
five-parameter NLO fit of the full photon structure
function in the $\disg$ scheme. It clearly falls from small to
large $x$ and $Q^2$ and amounts to only a few percent in the region that has
been used in the pointlike fit.

For our full (pointlike and hadronic) fit, we start from the observations that
$\f2y$ is dominated by the $u$-quark density in the photon and is only
sensitive to the combined density of $d$- and $s$-quarks, whose contribution is
furthermore suppressed by the smaller $d$- and $s$-quark charges. 
The gluon contributes to $\f2y$ in LO only through a rather weak coupling to
the quark singlet density in the evolution equations. A consecutive
fit of the $u$-quark, $d$- and $s$-quark, and gluon densities shows, that only
the first is well constrained by $\f2y$ data and that the fit does not
improve, when more degrees of freedom are added. Therefore we do not impose a
hadronic boundary condition for the gluon and assume, that the hadronic
fluctuations of the photon are insensitive to the quark charge, {\it i.e.} we
identify the hadronic boundary conditions for $u$-quarks and $d$- and
$s$-quarks at the starting scale $Q_0$. Together with $\alpha_s(m_Z)$ and
$Q_0$, we then fit the parameters $N$, $\alpha$, and $\beta$ of our ansatz
$f_{u,\,d+s}^\gamma (x,Q_0^2)=Nx^\alpha(1-x)^\beta$ to the
full data set described above. In the first three lines of Tab.\ \ref{tab:2}
%
\begin{table}
\caption{\label{tab:2}$Q_0$, $\chi^2$/DF,
         and $\alpha_s(m_Z)$ values obtained in LO and
         NLO in the $\ms$ and $\disg$ factorization schemes with a
         five-parameter fit of
         the hadronic  photon structure function $\f2y$. Also shown are the 
         results obtained without LEP data.}
\begin{ruledtabular}
\begin{tabular}{lllc}
       Scheme & $Q_0$/GeV &~$\chi^2/$DF& $\alpha_s(m_Z)$ \\
\hline
       LO     & $0.79\pm0.18$&121/129& $0.1475\pm0.0074$(ex)$^{+0.0141}_{-0.0072}$(th)\\
       $\ms$  & $0.83\pm0.09$&118/129& $0.1198\pm0.0028$(ex)$^{+0.0034}_{-0.0046}$(th)\\
       $\disg$& $0.85\pm0.09$&115/129& $0.1216\pm0.0028$(ex)$^{+0.0033}_{-0.0050}$(th)\\
\hline
       w/o LEP& $0.46\pm0.10$&~37/~38& $0.1147\pm0.0047$(ex)$^{+0.0282}_{-0.0033}$(th)\\
\end{tabular}
\end{ruledtabular}
\end{table}
%
we list the $Q_0$, $\chi^2$/DF, and $\alpha_s(m_Z)$ values obtained with this
five-parameter fit in LO and NLO. The starting scale $Q_0$
is perturbatively stable and is found to be close to the masses of the light
vector mesons $\rho$, $\omega$, and $\phi$ in contrast to earlier claims that
the perturbative evolution of $\f2y$ sets in only at rather high values of
$Q_0\sim 2$ GeV \cite{Gordon:1994mu}. The individual and total values of
$\chi^2$/DF lie again around unity or below, so that the fitted full photon
structure functions describe the full data set well within the experimental
uncertainties. Note that the $\chi^2$ value for the four TPC/2$\gamma$ points
at $Q^2=2.8$
GeV$^2$, which have not been used in the fits, is 18.0
and thus very large. The gluon density, generated with $f_g^\gamma(x,Q_0^2)=0$,
turns out to be in good agreement with recent
H1 dijet data \cite{Adloff:2000bs}. The experimental errors on
the values of $Q_0$ and $\alpha_s(m_Z)$ reflect an increase in $\chi^2$ by one
unit, when all other fit parameters are kept fixed. Due to the larger number
of data points in the full fit, the experimental error turns out
much smaller than in the pointlike fit. When the full fit is performed without
the LEP data (last line of Tab.\ \ref{tab:2}), the experimental error is
almost doubled, {\it i.e.} the impact of the LEP data is again impressive.
A fit to LEP data only leads to almost identical results as the full fit.
The theoretical error in LO and without the LEP data gets a large asymmetric
contribution from doubling the factorization scale, which is highly
correlated with an increase in the fitted value of $Q_0$ and which is
drastically reduced in the full NLO fit. Similar
results as those listed in Tab.\ \ref{tab:2} are obtained, when only $u$-quarks
are assigned a hadronic boundary condition.

\begin{figure}
 \centering
 \includegraphics[width=0.9\columnwidth]{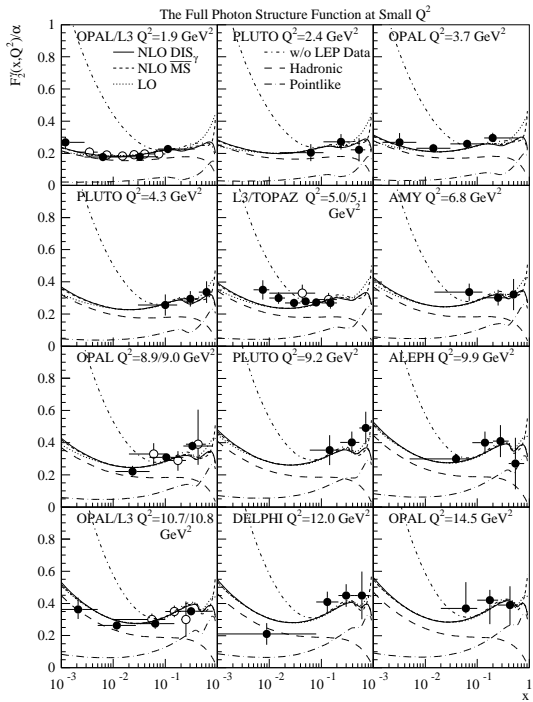}
 \caption{\label{fig:2}Five-parameter fits of the full photon structure
 function, compared to data from PETRA \cite{Berger:1984xt}, TRISTAN
 \cite{Kojima:1997wg,Muramatsu:1994rq}, and LEP \cite{Barate:1999qy,
 Abreu:1996xt,Acciarri:1998ig,
 Ackerstaff:1997ni,
 Abbiendi:2000cw} at small $Q^2$. The data points marked by open
 circles refer to the second experiment and/or $Q^2$ value. Also shown are the
 hadronic and pointlike contributions to the NLO fit in the $\disg$ scheme.}
\end{figure}

In Fig.\ \ref{fig:2}
%
%
we compare our results to the fitted $\f2y$ data in the region of low $x$
and $Q^2$. This region is clearly dominated by the hadronic contribution
and by the impact of the LEP data. A fit without the LEP data results in a rise
of $\f2y$ at low $x$, which is much too steep. The fits are perturbatively
stable and the data are described almost equally well in the $\ms$ and $\disg$
scheme.

Since the total error on $\alpha_s(m_Z)$ is smaller in the full fit than in
the pointlike fit due to the larger number of data points, we adopt as our
final result
\pagebreak
\beq
 \alpha_s(m_Z)=0.1198\pm0.0054
\eeq
in NLO and the $\ms$ scheme, where the larger theoretical error has been added
to the experimental error in quadrature. While our total error is
slightly larger than those obtained in $Z$-boson- and
$\tau$-decays at LEP, it is comparable to the errors obtained in deep-inelastic
scattering at HERA and heavy quarkonium decays. This encourages us to
combine our result with the current world average of $0.1172\pm0.0014$
\cite{Groom:2000in} to a new world average
\beq
 \alpha_s(m_Z)=0.1175\pm0.0014,
\eeq
where the errors are assumed to be uncorrelated.

In conclusion, we have for the first time fitted the now final PETRA,
TRISTAN, and LEP data on the photon structure function $\f2y$ in NLO of
perturbative QCD. We have extracted the value of the strong coupling constant
$\alpha_s(m_Z)$ with competitive experimental and theoretical errors from a
single-parameter pointlike fit to data at large $x$ and $Q^2$ and from a
five-parameter full (pointlike and hadronic) fit at all $x$ and $Q^2$.
Our analysis proves that the available $\f2y$ data contribute significantly
to a precise determination of $\alpha_s$ and that future measurements of
$\f2y$ at linear colliders will have a large impact.


\begin{acknowledgments}

We thank G.\ Kramer for many valuable discussions and a careful reading of the
manuscript. S.\ A.\ and M.\ K.\ are supported by the Deutsche
Forschungsgemeinschaft through Grant No.\ KL~1266/1-2. 

\end{acknowledgments}



\end{document}